\documentstyle[prl,twocolumn,aps,graphicx]{revtex}
\begin{document}
\draft

%----------------------User's Commands----------------------------
 \newcommand{\mytitle}[1]{
 \twocolumn[\hsize\textwidth\columnwidth\hsize
 \csname@twocolumnfalse\endcsname #1 \vspace{1mm}]}
%------------------------------------------------------------------------

\mytitle{
\title{Flux-Dependent Level Attraction in Double-Dot Aharonov-Bohm 
Interferometers}

\author{Bj\"orn Kubala and J\"urgen K\"onig}
\address{
Institut f\"ur Theoretische Festk\"orperphysik, Universit\"at Karlsruhe,
76128 Karlsruhe, Germany}

\date{\today}

\maketitle

\begin{abstract}

We study electron transport through a closed Aharonov-Bohm interferometer 
containing two noninteracting single-level quantum dots.
The quantum-dot levels are coupled to each other indirectly via the leads.
We find that this coupling yields signatures of an effective {\it flux-dependent 
level attraction} in the linear conductance.
Furthermore, we predict a {\it suppression of transport} when both levels are 
close to the Fermi level of the leads.
The width of this anomaly is also flux dependent.
We identify different regimes in which constructive interference of transmission 
through identical dots yields a signal that is 1, 2, or 4 times as large as the 
conductance through a single dot.

\end{abstract}
}

%
% 73.23.Hk Electronic transport in mesoscopic systems:
%          Coulomb blockade; single-electron tunneling
% 73.63.Kv Electronic transport in mesoscopic or nanoscale materials and
%          structures: Quantum dots
% 73.40.Gk Electronic transport in interface structures: Tunneling
%

{\it Introduction.} ---
The presence of quantum coherence in mesoscopic systems is detectable through 
interference experiments.
Transport measurements through multiply connected geometries containing a
quantum dot revealed oscillations of the conductance as a function of magnetic
flux, i.e., Aharonov-Bohm (AB) oscillations
\cite{Yacobi95,Schuster97,Ji00,Wiel00}, provided that the phase coherence length is larger than the dimensions of the device.
Differences between closed (two-terminal) AB interferometers 
\cite{Yacobi95,Wiel00,Yeyati95,Hackenbroich96,Bruder96,Baltin99,Entin01} and 
those with open geometries \cite{Schuster97,Entin01} have been discussed, and 
Kondo correlations \cite{Ji00,Gerland00,Hofstetter01}, the Fano effect
\cite{Hofstetter01,Entin01.2}, and the influence of Coulomb interaction on 
quantum coherence \cite{Koenig01,Koenig01.2} have been addressed.

More recently, an AB interferometer containing two quantum dots has been 
realized \cite{Holleitner00}.
The possibility to manipulate each of the quantum dots separately enlarges 
the dimension of the parameter space for the transport properties as compared to
a single-dot AB interferometer, and the enterprise to experimentally explore the
unknown territory has just begun.
Theoretical work on transport through double-dot systems includes studies of 
resonant tunneling \cite{Shahbazyan94} and 
cotunneling \cite{Akera93,Loss00}, as well as the prediction of asymmetric 
interference patterns \cite{Koenig01,Koenig01.2}, signatures of Kondo 
correlations \cite{Boese01}, and a quantum phase transition 
\cite{Hofstetter01.2} in the presence of strong Coulomb interactions.
Spectral properties of double dots coupled to leads have been studied 
in Ref.~\cite{KGS98}.

In this paper, we study a simple model system of a closed double-dot AB 
interferometer, which can be solved exactly.
Surprisingly, we find even for this model complex characteristic transport 
features such as signatures of a flux-dependent level attraction and an anomaly 
of suppressed transport, which can easily be manipulated by applied gate 
voltages and magnetic flux.
Our aim is to provide a map with the most significant transport signals, which 
may serve as a guide for the ongoing and future experimental endeavor.

{\it Model.} ---
We consider an AB geometry as depicted in Fig.~\ref{fig1}.
Two single-level quantum dots are coupled to leads, described by the standard 
tunnel Hamiltonian,
\begin{equation}
   H = \sum_{kr} \epsilon_{kr} a^\dagger_{kr}a_{kr} +
   \sum_{i=1,2} \epsilon_i c^\dagger_i c_i +
   \sum_{kri} (t_{ri} a^\dagger_{k r} c_i + {\rm H.c.}) \, ,
\label{Hamiltonian}
\end{equation}
where $a^\dagger_{kr}$ and $a_{kr}$ are the creation and annihilation operators 
for electrons with quantum number $k$ in the left or right lead, $r=L$ or $R$,
respectively, and $c^\dagger_i$ and $c_i$ are the Fermi operators for the states
in dot $i=1,2$.
The level energies in the dots (measured from the Fermi energy of the leads) are 
denoted by $\epsilon_1$ and $\epsilon_2$.
They can be varied by applied gate voltages.
It is convenient for the following calculations to define the average level 
energy $\bar \epsilon = (\epsilon_1 + \epsilon_2)/2$ and the difference
$\Delta \epsilon = \epsilon_2 - \epsilon_1$.
We neglect the energy dependence of the tunnel matrix elements $t_{ri}$ 
\cite{note_t} and assume a symmetric coupling strength $|t_{ri}| = |t|$.
Due to tunneling, each dot level acquires a finite linewidth
$\Gamma=\Gamma_{L}+\Gamma_{R}$ with $\Gamma_{r}=2\pi|t|^2N_{r}$,
where $N_r$ is the density of states in lead $r=L,R$.
The magnetic flux is modeled by an AB phase attached to the tunnel matrix 
elements \cite{model_phi}.
We choose a symmetric gauge such that
$(t_{L1})^* = t_{L2} = (t_{R2})^* = t_{R1} = |t| \exp(i\varphi/4)$, with 
$\varphi \equiv 2\pi \Phi/\Phi_0$, and $\Phi_0 = h/e$ is the flux quantum.

In the above model there is no direct interaction (either Coulomb repulsion or
tunnel coupling) between the two quantum dots.
The levels are rather coupled indirectly to each other via the leads.
Furthermore, since for each dot only one level supports the transport (for the 
level spacing being larger than bias voltage, linewidth, and temperature), no 
intradot Coulomb interaction terms enter the 
Hamiltonian \cite{note_interaction}.

The experimentally accessible quantity is the linear conductance 
$G^{\rm lin} = (\partial I / \partial V)|_{V=0}$, which is related to the 
transmission $T(\omega)$ for an electron with energy $\omega$ by 
\begin{equation}
\label{nonlinear_current}
   G^{\rm lin} = - {e^2\over h} \int d \omega \, T (\omega) f'(\omega)  \, ,
\end{equation}
where $f'(\omega)$ is the derivative of the Fermi-Dirac distribution function.

{\it Exact transmission formula.} ---
Since Eq.~(\ref{Hamiltonian}) describes a model of noninteracting electrons, 
the total transmission $T(\omega)$ can be expressed \cite{Meir92} as 
\begin{equation}
\label{transmission_general}
   T(\omega) = {\rm \bf tr} \left\{{\bf G}^{\rm a} (\omega) {\bf \Gamma}^R 
     {\bf G}^{\rm r} (\omega) {\bf \Gamma}^L \right\}  \, ,
\end{equation}
where ${\bf G}^{\rm r/a} (\omega)$ is the matrix of retarded/advanced dot Green's
functions, and ${\bf \Gamma}^{L/R}$ describes the coupling to the left/right 
lead.
The matrix elements for the retarded Green's functions are defined in time space
as $G_{ij}^{\rm r}(t) = -i \Theta (t) \langle \{ c_i(t), c_j^\dagger(0) \} 
\rangle$.
The 2 $\times$ 2 matrix structure accounts for the two quantum dots (we set 
$\hbar =1$ from now on). The tunnel coupling is described 
by ${\bf \Gamma}^{L}= (\Gamma/2) \left( \begin{tabular}{cc} 1 & 
$\exp(i \varphi/2)$ \\ $\exp(- i \varphi/2)$ & 1 \end{tabular} \right)$
and ${\bf \Gamma}^{R} = \left( {\bf \Gamma}^{L}\right)^*$.

We obtain the exact equilibrium Green's functions by employing an 
equation-of-motions approach.
Briefly, this method consists of using the time evolution
$i\partial_t c_i = [c_i,H]$ to relate the time derivative
$\partial_t G_{ij}^{\rm r}(t)$ to $G_{ij}^{\rm r}(t)$ and new Green's functions
involving one dot- and one lead-electron operator.
We repeat this for these newly generated Green's functions until we get a closed 
set of equations. Eventually, we obtain the solution
\begin{equation}
\label{Green's function}
   {\bf G}^{\rm r}(\omega) =
        \left( \begin{array}{cc}
        \omega - \epsilon_1 + i {\Gamma \over 2}
        & i {\Gamma \over 2} \cos {\varphi \over 2}
        \\
        i {\Gamma \over 2} \cos {\varphi \over 2}
        & \omega - \epsilon_2 + i {\Gamma \over 2}
        \end{array} \right) ^{-1}
\end{equation}
and for ${\bf G}^{\rm a}(\omega)$ the complex conjugate.
Inserting this result into Eq.~(\ref{transmission_general}) leads to the total
transmission
\begin{equation}
  T(\omega) = 
  {\Gamma^2 \left[ (\omega - \bar \epsilon)^2 
      \cos^2 {\varphi \over 2}
      + ({\Delta \epsilon \over 2})^2 \sin^2 { \varphi \over 2 } \right] \over
    \left[ (\omega - \bar \epsilon)^2 - ({\Delta \epsilon \over 2})^2 
      - \left(\Gamma \over 2\right)^2 
      \sin^2 {\varphi \over 2} \right]^2 + (\omega - \bar \epsilon)^2 
    \Gamma^2 }
\label{general}
\end{equation}
This is the central and most general result of this paper.

{\it Level attraction and suppressed transport.} ---
We analyze the transport as a function of the bare energy-level positions (or, 
equivalently, the gate voltages).
At low temperature, the linear conductance is just $e^2/h$ times the 
transmission $T(\omega = 0)$ of incoming electrons at the Fermi energy.
The latter is shown in Fig.~\ref{fig2} for finite magnetic flux
(we arbitrarily choose the value $\varphi = 2\pi /5$).
We find that there are two striking features: lines of full transmission 
$T=1$ and a sharp anomaly of suppressed transport around 
$\epsilon_1 = \epsilon_2 = 0$.

The lines of full transmission $T=1$ form hyperbolas 
$\bar \epsilon^2 - (\Delta \epsilon /2)^2 = -(\Gamma/2)^2 \sin^2 (\varphi/2)$;
see thick solid lines in Fig.~\ref{fig3}.
Intuitively, one could interpret the incidence of full transmission as resonance 
of {\it renormalized} dot levels with the Fermi level of the leads.
Starting from bare levels $\epsilon_1, \epsilon_2$, we find the renormalized 
level positions as the $\omega$ values that satisfy $T(\omega)=1$.
Following this picture, we find no renormalization as long as the bare level 
energies are well separated, $|\Delta \epsilon| \gg \Gamma |\sin(\varphi/2)|$, 
but an effective {\it flux-dependent level attraction} otherwise.
This leads to lines of $T(\omega = 0)=1$ in the $\epsilon_1 \cdot \epsilon_2 < 0$
region of Fig.~\ref{fig3} (as opposed to lines in the  
$\epsilon_1 \cdot \epsilon_2 > 0$ region, which would indicate level 
{\it repulsion}).
The strength of the level attraction depends on the AB phase.
The maximum is achieved for odd-integer values of $\varphi/\pi$ (see dashed lines
separating white and shaded regions in Fig.~\ref{fig3} for full transmission), 
and level attraction vanishes at $\varphi = 0$; see diagonals (dotted lines).
We remark that there is a subtlety in interpreting full transmission as a
resonance condition.
The dot structure probed by transport may differ from the real one obtained 
from direct spectroscopy since different linear combinations of 
the bare dot levels couple differently strong to the leads.
In fact, it has been shown \cite{KGS98} that in the absence of magnetic flux our 
model shows a level attraction with the real level positions being defined by 
the maximum of the spectral density.
In contrast, the level positions defined by the transport signal as 
discussed in this paper are not renormalized at $\varphi = 0$.

Around the point $\bar \epsilon = \Delta \epsilon = 0$ there is a sharp anomaly 
of suppressed transmission [with $T=0$ at $\bar \epsilon = \Delta \epsilon = 0$].
The width of this dip is $|\sin(\varphi /2)|$, as it is bound by the lines 
of full transmission on the $\Delta \epsilon/(2\Gamma)$ axis and by saddle 
points [with height $T =\cos^2(\varphi/2)$] on the $\bar \epsilon/\Gamma$ axis
(see Figs.~\ref{fig2} and \ref{fig3}). For the special case 
$\Delta \epsilon=0$ this dip was already found in \cite{Shahbazyan94}. 
We emphasize that both the anomaly of suppressed transmission and the effective 
level attraction are {\it not} captured by a first- or second-order perturbation
expansion in $\Gamma$.

{\it AB oscillations}. --- 
We now discuss the shape of the AB oscillations, i.e., oscillations of the 
transmission as a function of magnetic flux for fixed level positions 
$\epsilon_1$ and $\epsilon_2$ (see Fig.~\ref{fig4}).
Two features will be emphasized: the evolution of sharp peaks close to the 
anomaly of suppressed transmission and a maximum-to-minimum transition of the 
transmission around the AB phase $\varphi = \pi$.

Away from the anomaly in the center, the oscillations are sinusoidal 
(curves $a$, $b$, and $g$ in Fig.~\ref{fig4}).
When entering the region of the anomaly in the center of the diagram in 
Fig.~\ref{fig3}, higher-harmonic contributions become important (see curves $e$ 
and $f$). 
These correspond to paths through the AB geometry with higher winding number 
around the enclosed flux (the phase coherence length has to be longer than 
these paths).
Close to the center, sharp peaks around AB-phase values 
$0, \pm 2\pi, \pm 4\pi, \ldots$ result (curves $c$ and $d$). 
This opens the possibility to manipulate transport in a nontrivial way by 
varying the magnetic field. 
{\it The sensitivity of this dependence is determined by the gate voltages of 
the quantum dots.}

The behavior of the transmission near flux values $\pm \pi, \pm 3 \pi, \ldots$ 
underpins the notion of an effective flux dependent level attraction. 
In the regime indicated by the white region in Fig.~\ref{fig3}, the AB 
oscillations show a minimum as a function of $\varphi$, while in the shaded 
region a maximum occurs.
This is consistent with interpreting the lines of full transmission as the 
renormalized energy levels being in resonance with the Fermi level, as we did 
above:
In the shaded region, the two {\it renormalized} dot levels are on opposite 
sides of the Fermi energy while they are on the same side in the white region.
If a dot level is lying above the Fermi energy, particle-like 
processes will dominate transport through that dot, while hole-like processes 
dominate in the opposite case.
The corresponding transmission phases differ by $\pi$ which explains
the maximum to minimum transition.

{\it Fano line shapes.} ---
Interference between resonant transport through a single level and a continuous 
background channel yields asymmetric line shape for the conductance as a function
of the level position, the well-known Fano effect \cite{fanoorg}.
Within our model we can simulate such a situation by keeping one energy level, 
say $\epsilon_2$, fixed and varying the other one $\epsilon_1$.
Transport through quantum dot 2 provides then the ``background channel'' with 
transmission $T_{\rm b} = (\Gamma/2)^2/[\epsilon_2^2+(\Gamma/2)^2]$.
After defining $e\equiv - [\epsilon_1 + {\rm Re} \Sigma(0)] / {\rm Im} 
\Sigma(0)$ with $G^{\rm r}_{11}(\omega)=1/[\omega-\epsilon_1-\Sigma(\omega)]$ 
obtained by Eq.~(\ref{Green's function})
and the Fano parameter $q \equiv (2\epsilon_2/\Gamma) 
[-1+(2-T_{\rm b})\cos^2 (\varphi/2)]/[1-T_{\rm b} \cos^2(\varphi/2)]$, we find 
the generalized Fano form
\begin{equation}
  T = T_{\rm b} {(e+q)^2\over e^2+1} + {A \sin^2 \varphi \over e^2+1}
\label{fano}
\end{equation}
at $\omega=0$, with $A = (1-T_{\rm b}) / [1-T_{\rm b} \cos^2 (\varphi / 2)]^2$.
For dot level 2 tuned far away from resonance, $|\epsilon_2| \gg \Gamma/2$,
$A$ approaches unity \cite{note}.

{\it Destructive and constructive interference}. --- 
The textbook example for quantum interference effects is the two-slit experiment.
The standard way to demonstrate destructive and constructive interference is to
consider a setup where the moduli of the transmission amplitudes through either 
slit are identical, $|t_1| = |t_2| = |t|$, and to tune the relative 
phase such that the total transmission probability $|t_1 + t_2|^2$ becomes 
extremal, i.e., $0$ for destructive and $4|t|^2$ for constructive interference.
There is, however, a principal difference between this two-slit setup and
the double-dot AB interferometer we study.
In the former one only a fraction of the emitted particles reach the 
detector while most of them are scattered to the periphery.
The latter has a closed geometry, and all incoming electron must either arrive 
at the drain or be backscattered to the source.
Therefore, we can ask the question whether destructive and constructive 
interference will still emerge in our model system.

To make the analogy to the two-slit setup as close as possible, we consider 
equal level energies $\epsilon_1 = \epsilon_2 = \epsilon$.
It is easy to see from Eq.~(\ref{general}) that destructive interference,
$T=0$, is achieved for $\varphi$ being an odd multiple of $\pi$, which proves 
that in our model the transport is fully coherent for all temperatures and 
coupling strengths \cite{note2}.

The situation $\varphi = 0$ corresponds to the constructive-interference
scenario in the two-slit experiment.
At $\varphi=0$ (and $\Delta \epsilon =0$) the transmission through the
double-dot AB interferometer has Breit-Wigner form
$T_{\rm 2 dot}(\omega) = \Gamma^2 / [(\omega-\epsilon)^2 + \Gamma^2]$, but with 
a level width twice as large as for a single dot, 
$T_{\rm 1 dot}(\omega) = (\Gamma/2)^2 / [(\omega-\epsilon)^2 + (\Gamma/2)^2]$.
This can be easily understood by writing the Hamiltonian in terms of symmetric 
and antisymmetric combinations of the dot levels to see that the antisymmetric 
combination decouples, whereas the symmetric combination
acquires an increased coupling strength, $t\rightarrow \sqrt{2}t$.

It follows that at low temperature and at resonance 
$|\epsilon|, k_B T \ll \Gamma$, the linear conductance through the double-dot
system is equal to that through a single dot $G^{\rm lin}_{\rm 1 dot} =e^2/h$
in the absence of the other arm of the interferometer, 
${G^{\rm lin}_{\rm 2 dot} / G^{\rm lin}_{\rm 1 dot}} = 1$.

At high temperature, $|\epsilon|, \Gamma \ll k_B T$, the conductance is 
dominated by contributions in first order in $\Gamma$, and subsequently, we 
obtain ${G^{\rm lin}_{\rm 2 dot} / G^{\rm lin}_{\rm 1 dot}} = 2$ (see also
Refs.~\cite{Koenig01,Koenig01.2}).

It is only in the regime $\Gamma, k_B T \ll |\epsilon|$ (the so-called 
cotunneling regime, in which transport is of order $\Gamma^2$) that the ratio
${G^{\rm lin}_{\rm 2 dot} / G^{\rm lin}_{\rm 1 dot}} = 4$ reaches the value 
as for constructive interference in the two-slit experiment.

{\it Summary.} ---
We studied transport through an Aharonov-Bohm interferometer containing two 
noninteracting, single-level quantum dots.
Based on the derivation of an exact expression for the total transmission we 
found signatures of a flux-dependent level attraction and an anomaly of 
suppressed transport.
We analyzed the form of AB oscillations in different regions of the parameter
space, and found the evolution of sharp peaks near the anomaly of suppressed 
transport and a maximum-to-minimum transition of the AB signal around 
$\varphi = \pi$.
Regimes where constructive interference through identical dots yields a 
transmission that is 1, 2, or 4 as large as that through a single 
quantum dot were identified.

{\it Acknowledgments.} ---
We acknowledge helpful discussions with D. Boese, Y. Gefen, Y. Imry, 
H. Schoeller, and G. Sch\"on.
This work was supported by the Deutsche Forschungsgemeinschaft through the
Emmy-Noether program and the Center for Functional Nanostructures.

\begin{figure}
\centerline{\includegraphics[width=8cm]{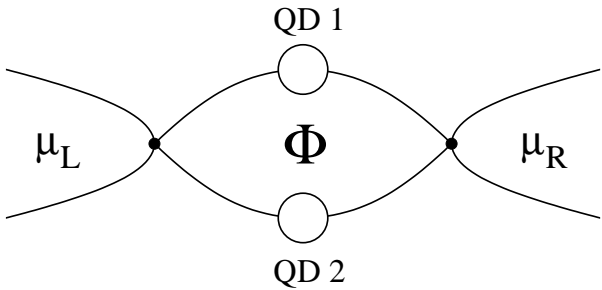}}
\caption{Double-Dot Aharonov-Bohm interferometer.}
\label{fig1}
\end{figure}
\begin{figure}
\centerline{\includegraphics[width=8cm]{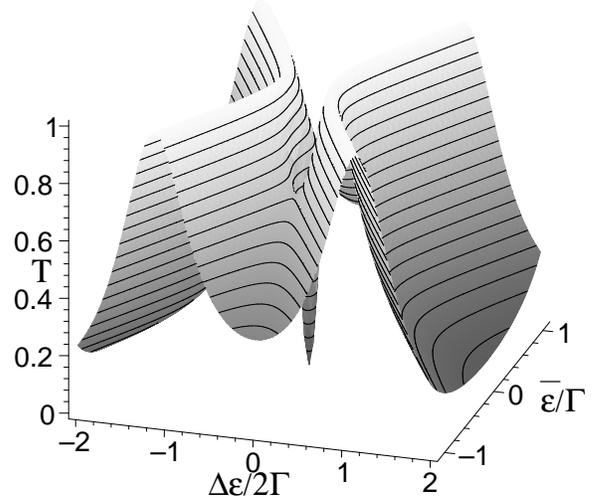}}
\caption{Transmission $T(\omega = 0)$ as a function of the average 
  $\bar \epsilon$ and difference $\Delta \epsilon$ of the dot level energies for
  $\varphi = 2\pi/5$.}
\label{fig2}
\end{figure}
\begin{figure}
\centerline{\includegraphics[height=8cm,angle=-90]{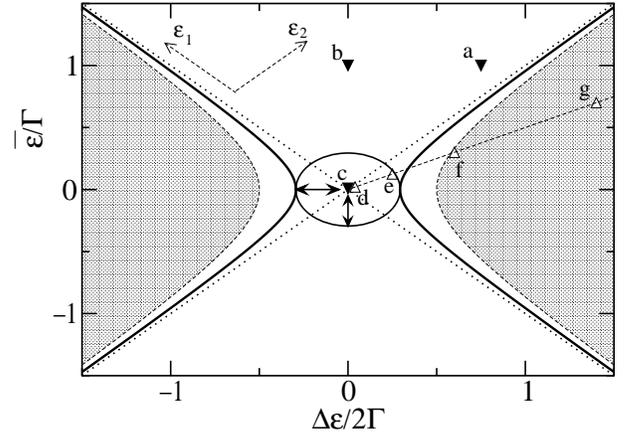}}
\vspace*{.4cm}
\caption{Sketch of the significant features of the transmission for  
  $\varphi = 2\pi/5$.
  Thick solid lines (hyperpolas) denote full transmission, $T=1$.
  Dotted lines (diagonals) indicate the lines of full transmission for the case
  $\varphi = 0$.
  The circle in the middle sketches the boundary of the anomaly of suppressed 
  transmission. 
  Its half width (arrows) is $|\sin (\varphi/2)|/2$.
  In the white and shaded regions, AB oscillations show a minimum and a maximum
  at flux $\varphi = \pi$, respectively (see Fig.~\ref{fig4}).}
\label{fig3}
\end{figure}
\begin{figure}
\centerline{\includegraphics[height=8cm,angle=-90]{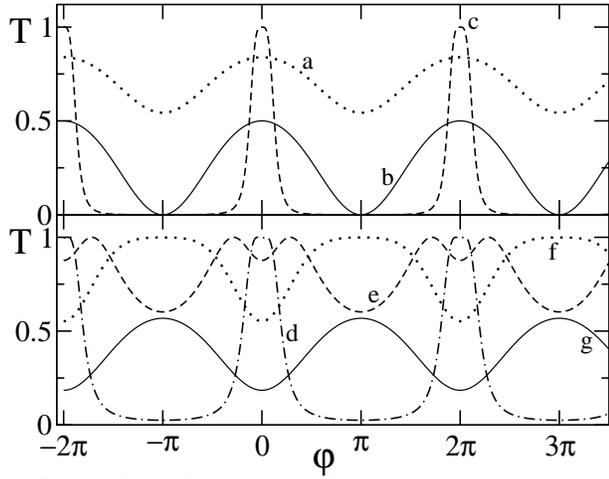}}
\caption{AB oscillations for different values of 
  $(\Delta \epsilon / 2\Gamma, \bar \epsilon / \Gamma)$ as indicated in 
  Fig.~\ref{fig3}.
  a: (0.75,1), b: (0,1), c: (0,0.01), d: (0.04,0.02), e: (0.25,0.125), 
  f: (0.6,0.3), and g: (1.4,0.7).}
\label{fig4}
\end{figure}


\begin{references}
\bibitem{note_t}
This widely used approximation works since only lead states close to the Fermi 
energy participate in transport.
\bibitem{model_phi}
A magnetic-field dependence of $\epsilon_1$ and $\epsilon_2$ is unimportant for
the subsequent discussion.
It would just shift the curves shown in Figs.~\ref{fig2} and \ref{fig3},
and the corrections to the AB oscillations shown in Fig.~\ref{fig4} would be
negligible as long as the area of each quantum dot is small as compared to the
area enclosed by the interferometer arms.
\bibitem{note_interaction}
Such terms are absent not only in devices, in which Coulumb interaction within 
each dot can be neglected, but also when spin degeneracy is lifted in a strong 
Zeemann field or in an all-ferromagnetic sample.
\bibitem{note}
We note that Eq.~(\ref{fano}) and the definition of the Fano 
parameter $q$ are identical to those for a closed single-dot AB interferometer 
(see Eqs.~(4) and (5) of Ref.~\cite{Hofstetter01}) but with a renormalized flux 
$\tilde \varphi$ defined by $\sin^2 \tilde \varphi = A \sin^2 \varphi$.
\bibitem{note2}
The apparent contradiction to a naive sequential-tunneling picture which yields
incoherent transport at high temperatures has been addressed in
Refs.~\cite{Koenig01,Koenig01.2}.
\end{references}
\end{document}